\documentclass[a4paper,UKenglish,cleveref, autoref, th-restate]{lipics-v2021}
\hideLIPIcs
\nolinenumbers

\usepackage{tikz}
\usetikzlibrary{decorations.pathmorphing,calc,positioning,arrows.meta}
\newcommand{\debug}[1]{}

\newcommand{\lgn}{\log n}
\newcommand{\lglgn}{\log \log n}
\newcommand{\poly}[1]{\mathrm{poly}(#1)}
\newcommand{\plg}[1]{\poly{\log #1}}
\newcommand{\plglg}[1]{\plg{\log #1}}

\newcommand{\dfs}[1]{\mathrm{num}(#1)}
\newcommand{\lp}[1]{\mathrm{lowpoint}(#1)}
\newcommand{\lca}[1]{\mathrm{lca}(#1)}
\newcommand{\nd}[1]{\mathrm{num\_descendants}(#1)}
\newcommand{\depth}[1]{\mathrm{depth}(#1)}
\newcommand{\children}[1]{\mathrm{children}(#1)}
\newcommand{\parent}[1]{\mathrm{parent}(#1)}

\newcommand{\ops}{
    \begin{itemize}
    \item $\children{u}$ iteration over all children of $u$.
    \item $\parent{u}$: output the parent of $u$.
    \item $\dfs{u}$: return the $s$-$t$ number of vertex $u$.
    \item $\nd{u}$: return the number of descendants of $u$.
    \item $\depth{u}$: return the depth of $u$.
    \item $\lp{u}$: return the lowpoint of $u$.
    \item $\lca{u,v}$: return the lowest-common ancestor of vertices $u$ and $v$.
    \end{itemize}
}

\title{Space-Efficient Depth-First Search via Augmented Succinct Graph
Encodings}
\author{Michael Elberfeld}{THM, University of Applied Sciences Mittelhessen,
Gie{\ss}en,
Germany}{michael.elberfeld@mni.th.de}{https://orcid.org/0000-0003-4179-7557}{}

\author{Frank Kammer}{THM, University of Applied Sciences Mittelhessen,
Gie{\ss}en,
Germany}{frank.kammer@mni.th.de}{https://orcid.org/0000-0002-2662-3471}{}

\author{Johannes Meintrup}{THM, University of Applied Sciences Mittelhessen,
Gie{\ss}en,
Germany}{johannes.meintrup@mni.th.de}{https://orcid.org/0000-0003-4001-1153}{}

\authorrunning{M. Elberfeld, F. Kammer and J. Meintrup}
\Copyright{Michael Elberfeld, Frank Kammer and Johannes Meintrup}



\EventEditors{John Q. Open and Joan R. Access}
\EventNoEds{2}
\EventLongTitle{42nd Conference on Very Important Topics (CVIT 2016)}
\EventShortTitle{CVIT 2016}
\EventAcronym{CVIT}
\EventYear{2016}
\EventDate{December 24--27, 2016}
\EventLocation{Little Whinging, United Kingdom}
\EventLogo{}
\SeriesVolume{42}
\ArticleNo{23}
\begin{document}

\setlength{\unitlength}{1mm}

\newcommand{\Newdot}{{\leavevmode\put(0,.63){\circle*{1.5}}}}

\maketitle

\begin{abstract}

  We call a graph $G$ separable if a balanced separator can be computed for $G$ of size
  $O(n^c)$ with $c<1$. Many real-world graphs are separable such as graphs of
  bounded genus, graphs of constant treewidth, and graphs excluding a fixed
  minor $H$. In particular, the well-known planar graphs are separable. We
  present a succinct encoding of separable graphs $G$ such that
  any number of depth-first searches DFS can be performed, from
  any given start vertex, each in
  $o(n)$ time with $o(n)$ additional bits. After the execution of a
  DFS, the succinct encoding of $G$ is augmented
  such that the DFS tree is encoded inside the encoding.
  Afterward, the encoding provides common DFS-related queries in constant time.
  These queries include queries such as lowest-common ancestor of two given
  vertices in the DFS tree or queries that output the lowpoint of a
  given vertex in the DFS tree. Furthermore, for planar graphs, we
  show that the succinct encoding can be computed in  $O(n)$ bits and expected
  linear time, and a compact variant can be constructed in $O(n)$ time and
  bits.

\end{abstract}
\setcounter{page}{0}
\newpage
\section{Introduction} \label{sec:intro}

Depth-first search (DFS) in graphs forms the backbone of algorithms
for a number of applications like finding vertex or edge cuts. A depth-first
search implicitly computes a tree (or forest), over the vertices of the graph
that is called a \emph{DFS tree} (or forest), which has crucial structural
properties that are commonly used for applications. Depth-first search that
runs in linear time utilizes two folklore data structures: a stack to keep
track of the current paths into the graph and an adjacency list to efficiently
iterate over the neighbors of a given vertex. Standard implementations of this
approach use $\Theta(n \log n)$ bits of space for the stack, $O(\log n)$ bits
for each vertex identifier where, throughout the paper, we use the standard
notation to refer with $n$ to the number of vertices of a given graph and with
$m$ to the number of its edges. Lowering the space requirements for depth-first
search to $O(n)$
bits while still maintaining a (nearly) linear runtime was the aim of a series
of works during the last years: one of the first algorithms is due to Asano et
al.~\cite{AsanoIKKOOSTU14} who reduced the space to $O(n)$ bits, but increased
the runtime to $O(m \log n)$. After several
improvements~\cite{BanerjeeCRS18,ElmasryHK15,ChakrabortyRS16} to this result,
Hagerup~\cite{Hagerup20} presented the current state-of-the-art algorithm using
$O(n)$ bits and $O(n + m \log^*n)$ time.

In many circumstances we do not need to be able to handle any possible input graph.
Instead, the input
graphs have a special structure that we can utilize to devise algorithms that
are more efficient than the ones handling the general case. Typical examples are
planar graphs which can be drawn in the plane without edge crossings, or
generalizations of it like graphs of bounded genus of graphs excluding a fixed
minor. What these graphs have in common is that they are sparse, meaning $m=O(n)$.
As there exists a DFS that uses $O(n+m)$ time and bits
on general graphs~\cite{BanerjeeCRS18,Hagerup20}, 
we can execute a DFS with $O(n)$ time and bits on sparse graphs.
If we shift our attention to sublinear-space DFS algorithms, we can
observe some strong indications from complexity, that no sublinear-space
polynomial time algorithm can exist for a special variant of DFS,
called \emph{lexicographical DFS}~\cite{AsanoIKKOOSTU14,ElmasryHK15}.
A concrete sublinear space algorithm for this problem restricted to planar graphs
has an unwieldy massive polynomial runtime~\cite{IzumiO20}, due to the reliance on the 
logspace reachability result of Reingold~\cite{Reingold08}.

As mentioned above, computing a DFS is often just the beginning or part
of an algorithm solving more involved graph problems like computing biconnected
and strongly-connected components. Typically, these applications of DFS
store certain meta information that is computed while traversing the
graph. 
Examples are \emph{$s$-$t$-numberings} that number the vertices in the
order of their exploration~\cite{Schmidt13}, and \emph{lowpoints}
that help to identify vertex and edge cuts~\cite{HopcroftT73}. If we want to
extend existing approaches for computing DFS that are efficient in both
space and time to this more general setting, we can not easily store the meta information
of a vertex
using any standard encoding. Again, this would use $O(\log n)$
bits for each vertex and, hence, creates a total memory
footprint of $\Omega(n \log n)$ bits. Algorithms working in our target time and
space regime store information of a similar kind to data structures that,
internally, use less space for each stored
word or devise new algorithmic approaches,
e.g., storing only partial data, such that the full data can be recovered with
additional computations, leading to a runtime and space
increase~\cite{ChakrabortyRS16,Hagerup20,KammerKL19}.

In this paper we present both (1) a data structure that is able to keep meta
information we compute for the vertices of a graph space-efficiently and (2) an
approach for computing DFS traversals space-efficiently as well as
extensions to some common applications. Our results are applicable to classes of
separable graphs which have balanced separator size $O(n^c)$ for some
$c<1$. This covers, in particular, planar graphs, graphs of bounded genus and
graphs excluding a fixed minor.

\subparagraph*{Data Structure: Nested Divisions and Augmentations.}

In this paper we present a succinct encoding for so-called separable graphs,
such as planar graphs, that provides the following functionalities.
After the encoding is initialized, a
DFS can be executed (from any given start vertex) in $o(n)$ time and
$o(n)$ additional bits and afterward information of the executed
DFS can be queried in constant time. 
If needed, a new DFS can be computed at any given time.
Such information includes the $s$-$t$-numbering of a vertex, and the
lowest-common ancestor of two vertices.
Furthermore, for planar graphs, we show that the succinct encoding can be
computed in expected linear time with $O(n)$ bits used during the
construction step, and a compact variant can be constructed in $O(n)$ time
and bits.

The data structure we present, called \emph{succinct nested division}, extends a
succinct encoding of separable graphs by Blelloch and
Farzen~\cite{BlellochF10}. The encoding of Blelloch and 
Farzan is built on dividing the input graph into smaller ``mini
pieces''. Mini pieces are, in turn, divided into even smaller ``micro pieces''
that are small enough such that relevant structural information about them can
be pre-computed and kept in a lookup table. The key property of this approach is
that there are only few ``boundary vertices'' in each piece that are contained
in multiple pieces. 
This suggests algorithms that, like the data structure itself, switch between
these three levels. In fact, different algorithm approaches are needed for
different levels to maintain our space and time bounds.

We augment the encoding of Blelloch and Farzan with additional data structures
and show that this can be used to design more complex queries than the standard
graph access queries provided by Blelloch and Farzan. Intuitively, the boundary
vertices act like relays--any interaction between two non-boundary vertices in
different pieces must necessarily pass through these relays. As such,
long-range dependencies (e.g., long paths), are efficiently mediated by the
sparse set of boundary vertices. Hence, pieces can often be considered in
isolation with few interactions between them, and all interactions between
vertices of different pieces are communicated via boundary vertices. To capture
this property we introduce values we call \emph{strongly local}. Roughly
speaking, a value is strongly local if, for a non-boundary vertex inside some
piece $P$, a query that returns the value can be evaluated only with a small
amount of information that is directly encoded in $P$ in addition to
information stored with the boundary vertices.
For our use-case of depths-first search, a value $\pi(u)$ of a
    vertex $u$ is strongly local if (1) for every ancestor $v$ of $u$, and every
    descendant $w$ of $u$ in the DFS tree it holds $|\pi(u) - \pi(v)| \leq
    O(k)$ and $|\pi(u) - \pi(w)| \leq O(\ell)$ where $k$ and $\ell$ are the
    respective distances from $u$ to $v$ and $u$ to $w$ in the DFS tree,
and (2)
we know the function to compute $\pi(v)$ from $\pi(u)$ and from $\pi(w)$.

\subparagraph*{Algorithms: Depths-First Search and Applications.}

Our succinct encoding for arbitrary separable graphs is constructed such that a
DFS can be performed directly on the encoding from any given starting
vertex. Afterward, various queries regarding the DFS traversal are
available. In particular, we directly provide the necessary queries that, e.g.,
Hopcroft and Tarjan's biconnected-component algorithm~\cite{HopcroftT73},
or Schmidt's algorithm for chain decompositions~\cite{Schmidt13}
requires. We effectively provide an interface that allows to implement
typical standard algorithms without the need to design specialized techniques.
The following theorem summarizes our main result. Note that the
runtime and bits are sublinear once the encoding is computed. The operations
presented in the next theorem are only a set of examples of queries that have
the previously outlined property of being strongly local, chosen such
that the previously mentioned standard algorithms can be executed directly.
Note that the following theorem applies only to connected graphs, but if a
given graph is not connected one can easily apply the following theorem
and the subsequent corollary to each connected component separately.

\newcommand{\thmain}{
  Let $G$ be a connected separable graph. There exists a succinct encoding $\mathcal{D}$
  of $G$
  that provides neighborhood iteration, adjacency and degree queries in constant time
  per element output.
  Additionally, it provides the operation $\mathrm{dfs}(u)$ that executes
  a DFS in $O(n/\plglg{n})=o(n)$ time and $o(n)$ additional bits from
  a given start vertex $u$ such that afterward the following queries are
  available in constant time for the last computed DFS tree.
  \ops{}
}

\begin{theorem}\label{th:main}
    \thmain{}
\end{theorem}

Replacing succinct with compact in Theorem~\ref{th:main}, and additionally
restricting $G$ to be a planar graph, the encoding can be constructed with
$O(n)$ bits and in $O(n)$ time. The change from succinct to compact is due to
the use of a certain succinct dictionary data structure of Raman et
al.~\cite{RamanRRS07} relying on hashing functions that can only be computed
in expected linear time. When we replace the dictionary with a simpler variant
of Baumann and Hagerup~\cite{BaumannH19}, we obtain a compact encoding that can
be computed much more easily. When we are fine with expected linear time, we
can use the dictionary of Raman et al., and thus can construct the succinct
encoding in $O(n)$ bits and expected $O(n)$ time.

For other graph classes the construction time and bits depend on the runtime of
the respective separator algorithms used as subroutines to divide the input
graph. Note that the information-theoretic lower bound for encoding separable
graphs is $cn + o(n)$ for some graph-class dependent constant $c$~\cite{BlellochF10}, and thus any
$O(n)$-bit representation of a separable graph is compact, but not necessarily
succinct.

\newcommand{\cormain}{
  Let $G$ be a connected planar graph. There exists a compact (succinct) encoding of $G$
  that can be constructed in $O(n)$ bits and $O(n)$ time (expected time)
  that provides the same functionalities as the encoding of
  Theorem~\ref{th:main}.
}
\begin{corollary}\label{cor:main}
    \cormain{}
\end{corollary}

\subparagraph*{Related Work on Graph Divisions.}

Recursive divisions have been used as the basis for algorithms and data
structures for decades, commonly based on the so-called $r$-division for planar
graphs, defined precisely in the next section. These divisions build on a
recursive application of the linear-time separator algorithm of Lipton and
Tarjan~\cite{LiptonT79}, and the improvement of Goodrich~\cite{Goodrich95} who
showed that the entire recursively application can be executed in linear
time---a standard approach would use $O(n \log n)$ time. This division technique
has been successfully applied to algorithmic problems such as maximum flow and
minimum cut and all-pairs shortest path~\cite{Federickson87}. Applications in
the field of data structures include decremental data structures for
connectivity~\cite{Holm18,HolmR24} where decremental refers to modifications of
a graph that remove vertices or edges.  None of the mentioned results have a
focus on space-efficiency, and therefore use other techniques internally
compared to our approach.

\subparagraph*{Structure of the Paper.} Section~\ref{sec:nested} details our first
contribution, the nested-division data structure and augmented variants of
it. Subsequently, in Section~\ref{sec:dfs}, we present the sublinear-space and
time execution of depths-first searches and related
applications, executed directly on our data structure.

\section{Nested Divisions}\label{sec:nested}

The present section contains an overview of what we view as a nested division,
beginning with the graph-theoretic properties they provide us with
(Subsection~\ref{ssec:gtpond}), and continue how Blelloch and Farzan used nested
divisions as a data structure to encode separable graphs
(Subsection~\ref{ssec:ndads}). Finally, we extend the section with 
our ideas of abstract
augmentations in nested divisions we provide to allow complex queries
(Subsection~\ref{ssec:and}). We use the notation $[i]$ for any integer $i$
to refer to the set $\{1, \ldots, i\}$.

\subsection{Graph-Theoretic Properties of Nested Divisions}\label{ssec:gtpond}

We describe a slightly generalized variant of the well-known concept of
$r$-divisions sketched in the introduction. We start with a small set of
definitions. A \emph{balanced separator} of a graph $G=(V, E)$ is a set of
vertices $S \subset V$ such that each connected component of $G[V \setminus S]$
contains at most a constant fraction of the vertices. The exact constant does
not matter to us. We call a class of graphs $\mathcal{G}$, closed under taking
vertex induced subgraphs, \emph{$O(n^{\epsilon})$-separable} if each $G=(V, E)\in \mathcal{G}$ has
a balanced separator $S \subset V$ of size $O(n^{\epsilon})$ for some constant
$0<\epsilon<1$ depending on $\mathcal{G}$. We simply say that a graph or class
of graphs is \emph{separable} if the size of the balanced separator is of no
particular concern to us.


Examples of separable graphs are planar graphs with balanced separators of size
$O(\sqrt{n})$, graphs of constant treewidth, graphs of bounded genus, and many
more. Note that it is well known that separable graphs have $O(n)$
edges~\cite{LiptonT79}. We define a variant of the well-known $r$-divisions as follows.

\begin{definition}[Relaxed Division]\label{def:relaxed}
  Let $G$ be a graph belong to an $O(n^{\epsilon})$-separable class of graphs
  for some $0 < \epsilon < 1$. An \emph{$(\alpha, r)$-relaxed division} is a
  decomposition of $G$ into a collection $\mathcal{P}$ of subgraphs, called
  \emph{pieces}, satisfying:
    \begin{enumerate}
    \item Each piece is a subgraph with at most $r$ vertices such that there are $\Theta(n/r)$ pieces in total.
    \item Every edge is assigned to exactly one piece (containing both endpoints of the edge).
    \item Each piece has $O(\alpha r^{\epsilon})$ \emph{boundary vertices}, which are
      vertices shared between different pieces.
    \end{enumerate}
    We call $\alpha$ the \emph{relaxation} and $r$ the \emph{piece size}.
    Edges between two boundary vertices are called \emph{boundary edges}, edges
    between two non-boundary vertices \emph{non-boundary edges} and all other
    edges \emph{transitional edges}.
\end{definition}

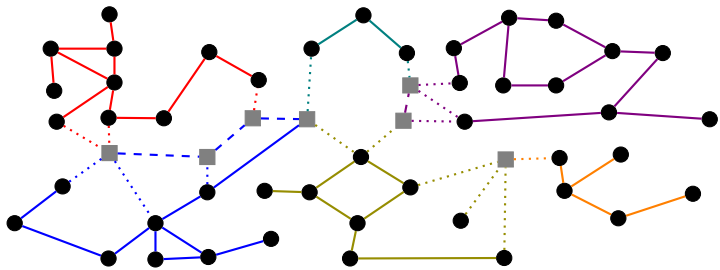
\begin{figure}[ht!]
    \centering
    \begin{tikzpicture}
    [simple/.style={circle, draw, minimum size=2mm, inner sep=0,fill=black},edge/.style={thick},
    boundary/.style={draw=gray, minimum size=2mm, inner sep=0,fill=gray},scale=0.65]
    \node[simple] (1) at (0.00, 0.74) {\debug{1}};
    \node[simple] (2) at (0.73, 4.30) {\debug{2}};
    \node[simple] (3) at (0.80, 3.44) {\debug{3}};
    \node[simple] (4) at (0.85, 2.81) {\debug{4}};
    \node[simple] (5) at (0.97, 1.49) {\debug{5}};
    \node[simple] (6) at (1.90, 0.02) {\debug{6}};
    \node[simple] (7) at (1.90, 2.89) {\debug{7}};
    \node[boundary] (8) at (1.92, 2.17) {\debug{8}};
    \node[simple] (9) at (1.92, 5.00) {\debug{9}};
    \node[simple] (10) at (2.02, 3.61) {\debug{10}};
    \node[simple] (11) at (2.02, 4.30) {\debug{11}};
    \node[simple] (12) at (2.85, 0.00) {\debug{12}};
    \node[simple] (13) at (2.85, 0.74) {\debug{13}};
    \node[simple] (14) at (3.02, 2.88) {\debug{14}};
    \node[simple] (15) at (3.90, 1.37) {\debug{15}};
    \node[boundary] (16) at (3.90, 2.09) {\debug{16}};
    \node[simple] (17) at (3.92, 0.05) {\debug{17}};
    \node[simple] (18) at (3.94, 4.23) {\debug{18}};
    \node[boundary] (19) at (4.82, 2.88) {\debug{19}};
    \node[simple] (20) at (4.94, 3.66) {\debug{20}};
    \node[simple] (21) at (5.06, 1.40) {\debug{21}};
    \node[simple] (22) at (5.19, 0.42) {\debug{22}};
    \node[boundary] (23) at (5.92, 2.86) {\debug{23}};
    \node[simple] (24) at (5.97, 1.37) {\debug{24}};
    \node[simple] (25) at (6.01, 4.30) {\debug{25}};
    \node[simple] (26) at (6.79, 0.02) {\debug{26}};
    \node[simple] (27) at (6.94, 0.74) {\debug{27}};
    \node[simple] (28) at (7.01, 2.09) {\debug{28}};
    \node[simple] (29) at (7.06, 4.98) {\debug{29}};
    \node[boundary] (30) at (7.87, 2.83) {\debug{30}};
    \node[simple] (31) at (7.94, 4.21) {\debug{31}};
    \node[simple] (32) at (8.01, 1.47) {\debug{32}};
    \node[boundary] (33) at (8.01, 3.55) {\debug{33}};
    \node[simple] (34) at (8.89, 4.31) {\debug{34}};
    \node[simple] (35) at (9.01, 3.60) {\debug{35}};
    \node[simple] (36) at (9.03, 0.79) {\debug{36}};
    \node[simple] (37) at (9.11, 2.81) {\debug{37}};
    \node[simple] (38) at (9.89, 3.55) {\debug{38}};
    \node[simple] (39) at (9.91, 0.03) {\debug{39}};
    \node[boundary] (40) at (9.94, 2.04) {\debug{40}};
    \node[simple] (41) at (10.01, 4.93) {\debug{41}};
    \node[simple] (42) at (10.96, 3.55) {\debug{42}};
    \node[simple] (43) at (10.96, 4.87) {\debug{43}};
    \node[simple] (44) at (11.03, 2.07) {\debug{44}};
    \node[simple] (45) at (11.13, 1.40) {\debug{45}};
    \node[simple] (46) at (12.10, 4.25) {\debug{46}};
    \node[simple] (47) at (12.22, 0.84) {\debug{47}};
    \node[simple] (48) at (12.27, 2.14) {\debug{48}};
    \node[simple] (49) at (12.03, 3) {\debug{49}};
    \node[simple] (50) at (13.12, 4.21) {\debug{50}};
    \node[simple] (51) at (13.73, 1.34) {\debug{51}};
    \node[simple] (52) at (14.07, 2.86) {\debug{52}};

    \draw[edge,blue] (1) -- (5);
    \draw[edge,blue] (1) -- (6);
    \draw[edge,red] (10) -- (11);
    \draw[edge,blue] (12) -- (13);
    \draw[edge,blue] (12) -- (17);
    \draw[edge,blue] (13) -- (15);
    \draw[edge,blue] (13) -- (17);
    \draw[edge,red] (14) -- (18);
    \draw[edge,blue,dotted] (15) -- (16);
    \draw[edge,blue] (15) -- (23);
    \draw[edge,blue,dashed] (16) -- (19);
    \draw[edge,blue] (17) -- (22);
    \draw[edge,red] (18) -- (20);
    \draw[edge,red,dotted] (19) -- (20);
    \draw[edge,blue,dashed] (19) -- (23);
    \draw[edge,red] (2) -- (10);
    \draw[edge,red] (2) -- (11);
    \draw[edge,red] (2) -- (3);
    \draw[edge,olive] (21) -- (24);
    \draw[edge,teal,dotted] (23) -- (25);
    \draw[edge,olive,dotted] (23) -- (28);
    \draw[edge,olive] (24) -- (27);
    \draw[edge,olive] (24) -- (28);
    \draw[edge,teal] (25) -- (29);
    \draw[edge,olive] (26) -- (27);
    \draw[edge,olive] (26) -- (39);
    \draw[edge,olive] (27) -- (32);
    \draw[edge,olive,dotted] (28) -- (30);
    \draw[edge,olive] (28) -- (32);
    \draw[edge,teal] (29) -- (31);
    \draw[edge,violet,dashed] (30) -- (33);
    \draw[edge,violet,dotted] (30) -- (37);
    \draw[edge,teal,dotted] (31) -- (33);
    \draw[edge,olive,dotted] (32) -- (40);
    \draw[edge,violet,dotted] (33) -- (35);
    \draw[edge,violet,dotted] (33) -- (37);
    \draw[edge,violet] (34) -- (35);
    \draw[edge,violet] (34) -- (41);
    \draw[edge,olive,dotted] (36) -- (40);
    \draw[edge,violet] (37) -- (49);
    \draw[edge,violet] (38) -- (41);
    \draw[edge,violet] (38) -- (42);
    \draw[edge,olive,dotted] (39) -- (40);
    \draw[edge,red] (4) -- (10);
    \draw[edge,red,dotted] (4) -- (8);
    \draw[edge,orange,dotted] (40) -- (44);
    \draw[edge,violet] (41) -- (43);
    \draw[edge,violet] (42) -- (46);
    \draw[edge,violet] (43) -- (46);
    \draw[edge,orange] (44) -- (45);
    \draw[edge,orange] (45) -- (47);
    \draw[edge,violet] (46) -- (50);
    \draw[edge,orange] (47) -- (51);
    \draw[edge,orange] (48) -- (45);
    \draw[edge,violet] (49) -- (50);
    \draw[edge,violet] (49) -- (52);
    \draw[edge,blue,dotted] (5) -- (8);
    \draw[edge,blue] (6) -- (13);
    \draw[edge,red] (7) -- (10);
    \draw[edge,red] (7) -- (14);
    \draw[edge,red,dotted] (7) -- (8);
    \draw[edge,blue,dotted] (8) -- (13);
    \draw[edge,blue,dashed] (8) -- (16);
    \draw[edge,red] (9) -- (11);
    \end{tikzpicture}
    \caption{
        A sketch of a division of graph into $6$ pieces of at most $13$
        vertices. The edges of each piece are colored with a distinct color.
        Boundary vertices are gray squares, and non-boundary vertices black
        circles. Non-boundary edges are solid, transitional edges dotted,
        and boundary edges dashed.
    }\label{fig:division}
\end{figure}

See Figure~\ref{fig:division} for a sketch of such a division.
The reason we introduce the relaxation $\alpha$, is due to our desire to
achieve space-efficiency when algorithmically constructing a division. There
exists an algorithm that computes a $(\lgn, r)$-relaxed division
via recursive separator searches that uses only $O(n)$ bits and, for
planar graphs, $O(n)$ time.  If the construction step of our encoding must not be
space-efficient, one can of course use a standard non-space efficient
algorithm~\cite{Goodrich95,KleinMSS13}, and therefore has a relaxation factor
of $\alpha=1$; for our application a factor of $\alpha=\log n$ makes no
difference. Common applications of divisions use them in a recursive fashion,
and so do we. In particular, we construct for each piece of some $(\alpha,
r)$-relaxed division a $(1, \tilde{r})$-relaxed division where $\tilde{r} \ll
r$. For this, we define a \emph{nested division} for a separable graph $G$
as follows.

\begin{definition}[Nested Division]
  Let $G$ be a graph belonging to an $O(n^{\epsilon})$-separable class of graphs
  for some $0 < \epsilon < 1$. An \emph{$(\alpha, r, \tilde{r})$-nested division} of
  $G$ is an $(\alpha, r)$-relaxed division of $G$ into pieces
  $\mathcal{P}=\{P_1, P_2, \ldots\}$ such that for each piece $P_i \in
  \mathcal{P}$ we have a $(1, \tilde{r})$-relaxed division $\mathcal{P}_i$.
    We call pieces of $\mathcal{P}$ \emph{mini pieces} and pieces of some
    $\mathcal{P}_i$ \emph{micro pieces}.
\end{definition}

If we have an $(\alpha, r, \tilde{r})$-nested division for a graph $G$ each vertex can
be viewed as being part of three \emph{levels}.
\begin{itemize}
    \item \emph{graph}: Each vertex is part of $V$. 
    \item\emph {mini}: Each vertex of $V$ is assigned to one or more mini pieces.
    \item\emph {micro}: Each vertex of $V$ is assigned to one or more micro pieces.
\end{itemize}

Given an $(\alpha, r, \tilde{r})$-nested division for a graph $G$ we can categorize all vertices as follows.
We call a vertex $\emph{mini (micro) boundary}$ if it is a boundary vertex in a mini (micro) piece,
and \emph{mini (micro) non-boundary} if it is not a boundary vertex in a mini (micro) piece.
We categorize edges similarly. We call an edge \emph{mini (micro) boundary edge}
if both endpoints are mini (micro) boundary vertices, \emph{mini} (micro) non-boundary edge
if both endpoints are mini (micro) non-boundary vertices, and \emph{mini (micro) transitional
edge} otherwise. We routinely drop the specifier mini or micro if we make statements that apply
to both mini and micro levels.

We are interested in bounding the number of total occurrences of
mini (micro) boundary vertices among all mini (micro) pieces, referred to as \emph{mini (micro)
duplicates}. Let $\mathcal{P}$ be an $(\alpha, r)$-relaxed division constructed
for an $O(n^{\epsilon})$-separable graph $G$. As each piece has
$O(r^{\epsilon})$ boundary vertices, the total number of duplicates is
$O((n/r) (\alpha r^{\epsilon}))$. Thus, for a nested division we have
$O(\alpha n/{r^{1-\epsilon}})$ (duplicates of) mini boundary vertices, and
$O((n/r)(r/\tilde{r})(\tilde{r}^{\epsilon}))=O(n/{\tilde{r}^{1-\epsilon}})$ (duplicates of) micro boundary vertices. Refer to a
more detailed description of the reasoning behind these bounds to Blelloch and
Farzan~\cite{BlellochF10}. We summarize this in the following lemma. 

\begin{lemma}[\cite{BlellochF10}]\label{lem:duplicates}
    Let $\mathcal{D}$ be a $(\alpha, r, \tilde{r})$-nested division constructed for an
    $O(n^{\epsilon})$-separable graph $G$ with relaxation parameters $\alpha$
    and $\alpha'$, and piece sizes $r$ and $\tilde{r}$. Then there are $O(\alpha
    n/{r^{1-\epsilon}})$ (duplicates of) mini boundary vertices and
    $O(n/{\tilde{r}^{1-\epsilon}})$ (duplicates of) micro boundary vertices.
\end{lemma}

For all our use cases of $(\alpha, r, \tilde{r})$-nested divisions we have
$\alpha=O(\log n)$, $r=\plg{n}$ and $\tilde{r}=\plglg{n}$, with the exact polynom
chosen such that we have $O(n/\plg{n})$
(duplicates of) mini boundary vertices, and $O(n/\plglg{n})$ (duplicates of)
micro boundary vertices. For the rest of this paper, when we refer to a nested
division we refer to a $(\log n, \plg{n}, \plglg{n})$-nested division.

\subsection{Nested Divisions as Data Structures}\label{ssec:ndads}

As we now describe concrete data structures, we fix our model of computation to
the word RAM model with word size $\Omega(\log n)$. This means we can perform constant-time bitwise operations
on binary values that occupy $O(\log n)$ bits. We begin by outlining
the most basic functionality we require for a nested division when used as a
data structure to encode a graph. This will give us the basis for subsequent
augmentations. Blelloch and Farzan presented a succinct encoding for separable
graphs that effectively uses a nested division at its core, and the framework
we present in this Subsection~\ref{ssec:ndads} is based on their work~\cite{BlellochF10},
defined slightly more general to allow for our later augmentations. While they
do not use an $\alpha$-relaxation factor for the initial $(\alpha, r)$-relaxed
division of the nested division, i.e., they use their own variation of a
standard recursively constructed $r$-division, their framework works with
$\alpha=\log n$ without any necessary modifications.
Their general idea
is to succinctly encode the nested division, and encode each piece at the micro level as
an index into a lookup table. Boundary vertices are encoded via additional data
structures that use standard non-space efficient techniques, e.g., using arrays
and lists. Combined with a succinct bidirectional mapping that maps a vertex to
its different occurrences in each level (graph, mini and micro) they realize
neighborhood, adjacency and degree queries. The mapping is constructed using
the powerful fully-indexable dictionary (FID) of Raman et
al.~\cite{RamanRRS07} that, for a given universe $[\ell]$, uses $o(\ell)$ bits
total when managing a set $S\subset[\ell]$ with $|S|=\Theta(\ell/\plg{\ell})$,
which fits Blelloch and Farzan's use-case.
Effectively, this FID manages a compressed bit vector $B$ of length $\ell$
where in $B$ bits at index $i \in S$ are set to $1$.
Due to the use of Raman et al.'s FID their construction takes
polynomial time $\omega(n)$ even for the simple case of encoding planar graphs
and without regard to space-efficiency during the construction step.
While Raman et al. mention their data structure can be constructed in expected
linear time, a deterministic linear time construction is not known. This is
due to the use of a certain hash function that is required.
When succinctness is not required much simpler dictionaries suffices for the
translation mappings. In this case, we effectively follow the same techniques
outlined by Blelloch and Farzan, but use the indexable dictionary of Baumann
and Hagerup, which can be constructed in $O(\ell/\log \ell)$ time and uses $\ell+o(\ell)$
bits for managing any set $S \subset [\ell]$.

\begin{lemma}[Fully Indexable Dictionary (FID)~\cite{BaumannH19}]\label{lem:fid}
    Let $B=(x_1, \ldots, x_{\ell})$ be a bit string. In $O(\ell/\log \ell)$ time and $\ell+o(\ell)$ bits
    a data structure can be constructed such that afterward the following
    queries can be executed in constant time for $b \in \{0, 1\}$.
    \begin{itemize}
        \item $\mathrm{rank}_{b}(i) = |\{j \in [i] | x_j = b\}|$
        \item $\mathrm{select}_{b}(i) = \mathrm{min}\{j \in [n] : \mathrm{rank}_B(j) = b\}$
    \end{itemize}
\end{lemma}

\subparagraph*{Translation Mapping.}
For the rest of this section, let $G$ be a separable graph and assume that a
nested division is given for $G$.
To describe the
functionality of the mappings, we introduce a labeling of pieces and vertices.
Each mini piece $P_i \in \mathcal{P}$ is uniquely identified by an id $i$.
Analogously, the relaxed division constructed for each
$P_i$ is identified as $\mathcal{P}_i$. Each micro piece $P_{i, j}$ is
identified by a tuple $(i, j)$ with $j$ indicating it
is the piece with id $j$, i.e., piece $P_{i, j} \in \mathcal{P}_i$. Each vertex $u \in
V$ has a \emph{graph label} assigned from $[n]$, a \emph{mini label $u_i$}
assigned from $[r]$ in each mini piece $P_i$ it is contained, and a \emph{micro
label} assigned from $[\tilde{r}]$ in each micro piece $P_{i, j}$ it is contained.
When we talk about a vertex $u \in V$ we always assume $u$ is identified by its
graph label. The mappings that are provided are as follows: given a vertex $u
\in V$ output the tuples $(i, u_i)$ such $u_i$ is the mini label of $u$ in the
piece $P_i$. Analogously for a given tuple $(i, u_i)$ where $i$ refers to a
mini piece $P_i$ and $u_i$ is the label of some vertex of $P_i$, output the
tuple $(j, u_{i, j})$ where $u_{i, j}$ is the micro label of vertex $u_i$ in a
micro piece $P_{i, j}$. Note that each of these queries can output more than
one element if the vertex is a boundary vertex at the respective level. For
non-boundary vertices, the query always returns a single element. Internally
these mappings are realized with a combination of standard data structures such
as lists and arrays for boundary vertices, as there are so few of them.
Additionally, a FID over a (compressed) bit vector $B$ of length $n$ is used
where a bit is set to $1$ at index $u$ exactly if $u$ is a mini boundary
vertex. Analogous FIDs over compressed bit vectors $B_i$ are constructed for
each mini piece $P_i$. This allows storing data for boundary vertices in
continuous memory
to store an array of $k$ entries
where each entry at index $i$ related to the boundary vertex
$B.\mathrm{select}_1(i)$. For less verbose writing we implicitly assume that we
always have access to all mappings outlined in this paragraph and only make
distinctions between the different types of labeling if necessary. Details can
be found in~\cite{BlellochF10}. We refer to the set of all outlined mappings as
\emph{translation mappings}.

\subparagraph*{Basic Graph Operations for each Level.}
Blelloch and Farzan showed how to implement a set of operations we refer to as
\emph{level graph queries}, all of which can be evaluated in constant time. The
operations include neighborhood iteration, degree queries and adjacency
queries. These operations can be executed on each level, with each respective
level requiring the following input parameters. On a graph level, the graph
label of a vertex suffices. On the mini level, we require the id of a mini
piece together with a mini label of a vertex, and analogously the id of a micro
piece together with a micro label of a vertex.

\subparagraph*{Bringing it all Together.} We end this section with a description of
a complete data structure, combining the previously described building blocks.
We refer to such a data structure built for $G$ with $\mathcal{D}_G$. If clear
from the context we drop the subscript and simply write $\mathcal{D}$. Recall
that the parameters of the nested division we construct are $\alpha=\log n$,
$r=\plg{n}$ and $\tilde{r}=\plglg{n}$.\footnote{Blelloch and Farzan use $\tilde{r}=\log n/\log
\log n$, but mention that other values are possible.}

\begin{lemma}[Succinct Nested Division,\cite{BlellochF10}]\label{lem:snd}
    Let $G \in \mathcal{G}$ be a graph and $\mathcal{G}$ a separable class of
    graphs. There exists a succinct data structure
    $\mathcal{D}$ of $G$ that represents a nested division of $G$ such that
    level mapping queries and level graph queries are provided in constant
    time.
\end{lemma}

For planar graphs we can construct the encoding efficiently, both in time and
space. Kammer and Meintrup presented a space-efficient algorithm for computing
a $(\log n, r)$-relaxed divisions of minor closed graph classes that roughly
works as follows. The input graph $G$ is replaced with a minor $F$ that
contains $O(n/\log n)$ vertices such that each vertex of $F$ represents
$\Theta(\log n)$ vertices of $G$. Then, any algorithm $\mathcal{A}$ can be used
for computing an $(1, r)$-relaxed division (a standard $r$-division) of $F$, which then can be turned
into a $(\log n, r)$-relaxed division of $G$. As $\mathcal{A}$ is executed
on a graph with $O(n/\log n)$ vertices, we achieve a speedup of a factor of
$\Theta(\log n)$, and a space reduction by the same factor. Translation between
the graph $F$ and $G$ uses linear time and bits.

When recursively constructing divisions of mini pieces we can use standard
algorithms as the graphs are small enough. For planar graphs, linear time
$r$-division algorithms~\cite{Goodrich95,KleinMSS13} exist, and thus the entire
computation of the recursive division runs in linear time. For other graph
classes the runtime depends on how efficiently separators can be found.
Using the FID of Lemma~\ref{lem:fid} (as outlined at the beginning of
Subsection~\ref{ssec:ndads}), we are able to construct a compact variant of the
encoding of Lemma~\ref{lem:snd} in $O(n)$ time and bits. If one is fine with
expected linear time, the FID of Raman et al.~\cite{RamanRRS07} can be used and
the encoding remains succinct. The functionalities of the different variants
are the same, and as such we make no distinction between them in the follow
sections and simply assume some nested division is given.

\begin{corollary}[Compact Nested Division]\label{lem:cnd}
   For planar graphs, a compact (succinct) variant of the encoding of
   Lemma~\ref{lem:snd} can be constructed with $O(n)$ bits and $O(n)$ time (expected time).
\end{corollary}

\subsection{Augmenting Nested Divisions}\label{ssec:and}

In the following we want to augment the nested division of
Lemma~\ref{lem:snd} with various capabilities. For this we analyze the
runtime and memory budget we can afford when our goal is to run some algorithms
in linear time and with linear bits, following by outlining necessary
techniques.

\subparagraph*{Memory and Runtime Budget.}
As stated, our goal is to maintain a space usage of $o(n)$ bits and a runtime
of $o(n)$ throughout our various applications of nested divisions, and we
construct our nested divisions with $r=\plg{n}$, $\tilde{r}=\plglg{n}$ and $\alpha =
\log n$. This means that for each of the $O(n/\plg{n})$ mini boundary
vertices we have a memory budget of $\plg{n}$ bits. Analogously, we have a
budget of $\plglg{n}$ bits for each of the $O(n/\plglg{n})$
micro-boundary vertices. This means, we have a budget of $\plg{n}$ bits for each
mini piece and $\plglg{n}$ for each micro piece. These bounds are actually far
more lax that what is needed for our applications, as we only require $O(\lgn)$
bits per mini boundary vertex (and mini piece), and $O(\lglgn)$ bits per micro
boundary vertex (and micro piece) for our applications. We have the same budget
for the runtime as we do for memory, but for our application only require
(amortized) constant time per boundary vertex.
An example for a data structure that fits within these bounds is an array with
an entry for each mini boundary vertex consisting of a constant number integers
in the range $[n]$, or an analogous array constructed
for a mini graph $P_i$ that contains an entry for each micro boundary vertex of $P_i$
such that the entry consists of a constant number of integers in the range
$[\lceil \lgn \rceil]$.

\subparagraph*{Table Lookup for Micro Pieces.} We have seen that nested divisions
are quite forgiving, and we can be generous with the amount of bits and time we
spend. For the next part, we have to be slightly more precise in our
calculations. In the following we denote with $\mathcal{Z}(n)$ the
information-theoretic lowerbound on the number of bits required to encode
graphs with $n$ vertices of the given separable graph class $\mathcal{G}$
for which we construct our encoding. We
store each micro piece as an index into a lookup table that lists all graphs
$G'$ of the given graph class $\mathcal{G}$ with at most $r^*\leq \tilde{r}$ vertices.
When the lookup table is constructed in such a way that no two listed graphs
are isomorphic, then each index requires $\mathcal{Z}(r^*) + O(1)$ bits.
The addition
of $O(1)$ is due to rounding to the nearest integer values. For separable
graphs $\mathcal{Z}(\cdot)$ is a linear function, and thus storing all micro
pieces as indices into this lookup tables can be done with $\mathcal{Z}(n) +
o(n)=O(n)$ total bits~\cite{BlellochF10}. Due to our choice of $\tilde{r}$ the size of the
lookup table is $\plg{n}$ bits, which is negligible. Additionally, the tiny
size of the lookup table lets us precompute and store any query we desire as
long as it can be computed and stored in $\mathrm{poly}(\tilde{r})$ time and bits (per
index). Any such query can be computed and output in $O(1)$ time in the word RAM model.
Examples are adjacency queries, neighborhood iteration or outputting a shortest
path between two given vertices, or computing a DFS.
Blelloch and Farzan used such a table lookup to store each micro piece of their
succinct encoding~\cite{BlellochF10}.

For our use-case these queries above are not enough. To give an example of why
this is not enough, consider a DFS. We are able to precompute a \textsc{DFS} (from any
given start vertex) for each of the tiny graphs $G'$ of the lookup table, such
that we can later output the entire DFS tree in constant time.
Later we run a DFS in the entire graph $G$, and as such certain information of
this global DFS must be encoded inside the piece represented by some graph
$G'$. Thus, we are not able to view a piece in total isolation which, so far,
is all that the lookup table provides us with.
Instead, we require the
graphs of the lookup table to be \emph{partially augmented}, meaning for at
most $b=O({\tilde{r}}^{\epsilon})$ vertices of each graph $G'$ of the lookup
table we store a bit string (called a \emph{partial vertex augmentation}) of some fixed
length $\ell=O(\log \tilde{r})$, and an additional bitstring of length $O(b
\ell)$ (called a \emph{partial graph augmentation}).
In our use-case, the $b$ augmented vertices will be the micro boundary vertices
inside a micro piece $P$, plus a constant number of micro non-boundary vertices. To
give a concrete example what we are able to store within these restrictions, we
can color the boundary vertices such that each of the colored vertices has a
unique color $c$, in addition to a sorted list $L$ of pairs of colors, such
that the entire list contains $O(b)$ such pairs. Each such pair $(c, c')$
can, for example, encode that a DFS has entered a piece $P$ via a boundary
vertex colored with $c$, and left it via a boundary vertex colored $c'$. 
We later use this scheme to reconstruct how a DFS has traversed a piece. 

As we want to have a succinct nested division, we can not store even a constant
number of colors for all vertices of a graph $G'$ of the lookup table. Instead,
we are able to compute a fixed coloring that depends only on $G'$, the partial vertex
augmentations of $G'$ (e.g., colored boundary vertices), and the partial graph
augmentation of $G'$ (e.g., a list with pairs of colors storing already used
path through the piece).
Our lookup table lists: all non-isomorphic graphs $G'$ of the given
graph class $\mathcal{G}$ together with all possible values for the partial vertex-
and graph
augmentation. Note that a single graph occurs many times with different values
of the augmentation, but no two 
graphs of the lookup table are isomorphic. The storage and computation of
the table can be done in $\plg{n}$ time and bits. The bit strings increases the
required bits per index by an additional term of $\log {\tilde{r} \choose
b}=O(b \log \tilde{r})$, which is $o(\log \log n)$ and therefore negligible. We
assume that we always have access to such a lookup table in the following sections.

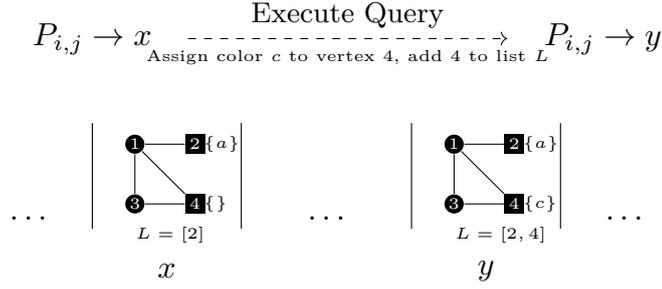
\begin{figure}[ht!]
    \centering

    \begin{tikzpicture}[
    simple/.style={circle, fill=black, inner sep=0pt, minimum size=6pt}, 
    every node/.style={scale=1.2}, 
    scale=1.4 
]

\begin{scope}[node distance=0.66cm] 
\node[simple] (A) {\tiny{\color{white}$1$}};
\node[simple, rectangle, right of=A] (B) {\tiny{\color{white}$2$}};
\node[rectangle,right=0.1cm of B.west] {\tiny{$\{a\}$}};
\node[simple, below of=A] (C) {\tiny{\color{white}$3$}};
\node[simple,rectangle,below of=B] (D) {\tiny{\color{white}$4$}};
\node[rectangle,right=0.1cm of D.west] {\tiny{$\{\}$}};
\draw[line width=0.4pt] (A) -- (B);
\draw[line width=0.4pt] (A) -- (C);
\draw[line width=0.4pt] (C) -- (D);
\draw[line width=0.4pt] (A) -- (D);
\draw[line width=0.4pt] (1,0.2) -> (1,-0.8);
\draw[line width=0.4pt] (-.4,0.2) -> (-.4,-0.8);

\node[rectangle, below right=0.15cm and 0cm of C.west] {\tiny{$L=[2]$}};
\end{scope}

\begin{scope}[xshift=3cm,node distance=0.66cm] 
\node[simple] (A) {\tiny{\color{white}$1$}};
\node[simple, rectangle ,right of=A] (B) {\tiny{\color{white}$2$}};
\node[rectangle,right=0.1cm of B.west] {\tiny{\{$a$\}}};
\node[simple, below of=A] (C) {\tiny{\color{white}$3$}};
\node[simple, rectangle, below of=B] (D) {\tiny{\color{white}$4$}};
\node[rectangle,right=0.1cm of D.west] {\tiny{$\{c\}$}};
\draw[line width=0.4pt] (A) -- (B);
\draw[line width=0.4pt] (A) -- (C);
\draw[line width=0.4pt] (C) -- (D);
\draw[line width=0.4pt] (A) -- (D);
\draw[line width=0.4pt] (1,0.2) -> (1,-0.8);
\draw[line width=0.4pt] (-.4,0.2) -> (-.4,-0.8);

\node[rectangle, below right=0.15cm and 0cm of C.west] {\tiny{$L=[2,4]$}};
\end{scope}

\node at (-1,-0.7) {$\ldots$};
\node at (1.8,-0.7) {$\ldots$};
\node at (4.6,-0.7) {$\ldots$};
\node at (0.3,-1.2) {$x$};
\node at (3.3,-1.2) {$y$};
\node at (-0.4,1) {$P_{i, j} \rightarrow x$};
\node at (4.4,1) {$P_{i, j} \rightarrow y$};

\draw[->,dashed] (0.5, 1) -- (3.5,1) node[midway,below,sloped] {\tiny{Assign color $c$ to vertex $4$, add $4$ to list $L$}}
        node[midway,above,sloped] {\small{Execute Query}};
\end{tikzpicture}

\caption{
    Initially the micro piece $P_{i, j}$ is stored as an index $x$ of the table
    lookup, sketched below with two entries explicitly shown. The graph shown
    in the table has regular vertices shown as circles, and the vertices for
    which we can store binary strings shown as rectangles, e.g., micro boundary
    vertices in the piece $P_{i, j}$.  Additionally, we have a list $L$ for each
    graph of the table that has enough capacity to store some values. 
    We execute a query that takes as input the index $x$, the vertex label
    $4$ and color $c$ to assign to vertex $4$. It returns the index $y$ of the lookup table such that
    $y$ stores the state that represents the execution of this query when
    applied to the graph stored at index $x$.  We then swap the index $x$ stored for
    $P_{i, j}$ with the index $y$ in our succinct nested division. 
}\label{fig:swap}
\end{figure}

\subparagraph*{Updating Indices.} The final ingredient of our algorithm is the ability to
\emph{swap indices}, which is described next. As mentioned, each micro piece is encoded
as an index into the lookup table referencing some graph $G'$ together with
its partial augmentations. When we want to update values stored
for (the micro boundary vertices of) the micro piece
we concretely do this by changing the index we store. Let $P_{i, j}$
be some micro piece encoded as an index $x$ into the lookup table. To update an
augmented value in the micro piece, e.g., color a
boundary vertex, we can replace the index $x$ with a different index $y$. See
Figure~\ref{fig:swap} for an example. 
We have to take care while constructing the table that any changes of the
partial augmentation (i.e., the values stored for boundary vertices) does not
change the internal labels of the graph, i.e., the names of the vertices stay
the same independent of the changes we make to the data stored for boundary
vertices. 

\section{Depths-First Search in Sublinear Time and Space}\label{sec:dfs}

We present our techniques for enabling the depths-first search directly on the
succinct encoding.  First, we show how the DFS is computed such that a so-called
palm tree is encoded inside the nested division
(Subsection~\ref{ssec:iterator}), followed by our description of how to provide
queries for so-called \emph{strongly local} values
(Subsection~\ref{ssec:meta}). As examples for strongly local values we describe
the technical realization of providing queries regarding the meta-information
of a DFS, e.g., $s$-$t$-numbering, lowpoints, lowest-common ancestors, and more.
We end the section with a description of how our main results can be realized
(Subsection~\ref{ssec:result}).

For this section, let $G=(V, E)$ be a connected separable graph given as a nested division
$\mathcal{D}$. We begin with the description of how to execute a depths-first
search (DFS) to compute a \emph{palm tree $T$} that is a directed version of
$G$ such that the edges $E$ are directed according to the DFS traversal, and by
their direction partitioned into two sets, tree edges and back edges.
Effectively, a palm tree is a DFS tree with additional 
back edges. Each vertex $u \in V(G)$ is assigned a number $\dfs{u}$
indicating at which point in time it was traversed, also known as an $s$-$t$-numbering. All edges $(u, v)$ of the palm tree are directed such that,
for a tree edge $\{u, v\}$ of the DFS tree, it holds that $\dfs{u} <
\dfs{v}$, and for all other edges $\{u, v\}$ (the back edges), it holds that
$\dfs{v} < \dfs{u}$.

In the next subsection we compute some palm tree $T$ directly on $\mathcal{D}$ and
afterward provide constant time queries regarding $T$ such as iteration over
all children of a vertex $u$ (i.e., all directed tree edges $(u, v)$). If we
are only interested in the directed tree edges of $T$, we refer to it as the
DFS tree.

\subsection{Iterator based DFS to compute a Palm Tree}\label{ssec:iterator}
In addition to the standard stack-based approach for implementing a
DFS, there is also an iterator approach that stores for each vertex an
iterator of the adjacency list. We implement the second variant.

\begin{figure}
    \centering
    \includegraphics[width=0.9\textwidth]{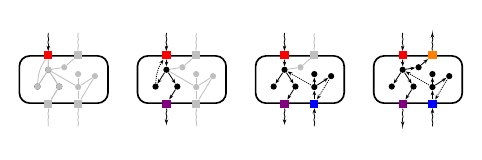}

    \caption{
        From left to right, the figure shows four DFS states of some piece $P$.
        In the first state, no vertex inside $P$ is visited yet, and we have
        just arrived at a boundary vertex of $P$ (colored red). We then advance
        the DFS until the next boundary vertex is visited, colored violet. 
        For micro pieces, we advance it via one query to the lookup table.
        We
        later arrive at the same piece again, this time via the blue boundary
        vertex. We enter the piece, and encounter the situation of the
        null-exit, thus we fully explore all vertices belonging to the subtree
        rooted at the blue vertex. Finally, we backtrack to the piece $P$ via
        the violet vertex, and see that there is a part of the piece not yet
        visited. We advance the state until the orange boundary vertex is
        visited and leave the piece again.
        The entry-exit pairs of the piece are $(\ {\color{red}\Newdot}\ ,%
        \ {\color{violet}\Newdot}\ )$, $(\ {\color{blue}\Newdot}\ ,\ {\texttt{null}})$ and
    $(\ {\color{red}\Newdot}\ ,\ {\color{orange}\Newdot}\ )$.
    Tree edges are shown as solid lines, and back edges as dotted lines.
    }\label{fig:dfsstate}
\end{figure}

Let $T$ be the palm tree constructed by some DFS traversal. See
Figure~\ref{fig:dfsstate} for an example of the following description. We say
$T$ \emph{enters} a piece $P$ if the there exists tree edges $(u, v)$ and $(v,
w)$ with only the edge $\{v, w\}$ part of the piece $P$, or if edge $\{v, w\}$
is part of the piece $P$, $(v, w)$ is a tree edge, and $v$ is the root of $T$.
We then call $v$ an \emph{entry vertex of $P$}. We say $T$ \emph{exits} $P$ if
there exist tree edges $(u,v)$ and $(v, w)$ with only $\{u, v\}$ part of $P$
and $\{v, w\}$ not part of $P$. We then call $v$ an \emph{exit vertex of $P$}.
We say $T$ \emph{starts} at a piece $P$ if the tree edge $(u, v)$ exists with
$u$ the root of $T$, $v$ is the first vertex visited after $u$ and $\{u, v\}$
is part of $P$. Note that the entry of a piece is by definition a boundary
vertex of some piece $P$, or the root of $T$. For us there is no difference
between the two cases of entry vertex (i.e., boundary vertex or root). We refer
to an exit vertex together with its matching entry vertex as an
\emph{entry-exit pair}. For easier description we refer to an entry vertex that
has no matching exit vertex as part of an entry-exit pair where the entry is
mapped to some special symbol, called the $\emph{null exit}$. E.g., the blue
vertex of Figure~\ref{fig:dfsstate} has no matching exit vertex. The following
observation directly follows. 

\begin{observation}\label{obs:exits}
    Let $k$ be the number of (micro/mini) boundary vertices contained in a
    (micro/mini) piece $P$ and $T$ any palm tree. Then there are $O(k)$
    entry-exit pairs associated with $P$.
\end{observation}

Let $P$ be some micro piece stored as an index to the lookup table. We define a
\emph{partial DFS state of $P$} as a coloring of the vertices of $P$ with the colors
unvisited (white), currently visited (gray) and finished (black), and a
(possibly empty) ordered set of entry-exit pairs,
ordered by the $s$-$t$-numbering of their respective entry vertex, with ties being
broken arbitrarily. 

 By the capabilities of the table lookup outlined in the previous section, an
 index of the lookup table is able to encode this information. The idea is
 that, when the DFS enters a micro piece $P$, categorized by some index $x$ of
 the lookup table encoding $P$ together with its partial DFS state, we obtain
 in $O(1)$ time an index $y$ of the lookup table encoding $P$ together with the
 next partial DFS state. We then swap out the index $x$ with the index $y$.
 Note that this query depends on: the piece $P$, the coloring of the vertices
 of $P$, the ordered set of entry-exit pairs and the vertex we used to just
 enter the piece $P$ (or the vertex we just backtracked to). This information
 is enough to obtain a new DFS state that correctly advances the \textsc{DFS} through
 $P$ while considering the current partial DFS state. Since such an advancement
 results in the DFS traversing up to (or backtracking to) the
 next micro boundary vertex we obtain a new entry-exit pair in $O(1)$ time.
 There are multiple possible viable partial DFS states that one could obtain by
 this operation, but we store one fixed state per entry of the lookup table,
 which does not matter. See Figure~\ref{fig:dfsstate} again for a visualization of
 how the DFS progresses through a piece. The change from one state of the
 figure to the next is done in $O(1)$ time for micro pieces.

A final note considers marking vertices as gray or black in micro pieces. When we
visit a micro boundary vertex $u$, we must color $u$ in all micro
pieces $P_{i, j}$ that contain $u$ (as micro label $u_{i, j}$). The translation
mappings allow us to iterate over all micro pieces that contain $u_{i, j}$ in
constant time per element output. For each such element output we must switch
out the respective index of the micro pieces via the lookup table. 
This is done exactly twice (white to gray, and gray to black) per duplicate of
a micro boundary vertices, and thus the time is linear in the number of total
micro boundary vertices.

We now describe the data structures required for implementing the DFS. We begin
with a description of how to implement an iterator over the neighborhood that
can be paused and resumed. Note that while the nested division $\mathcal{D}$
provides neighborhood iteration (Lemma~\ref{lem:snd}), it is not assumed that
this iteration can be paused and continued at a later point in time without
starting the process from the beginning. First, we describe this process for
micro boundary vertices $u_i$ in some piece $P_i$. For each such $u_i$ store an
array $A$ containing the indices of each micro piece $P_{i, j}$ that contains
$u_i$ (as micro label $u_{i, j}$). An iterator for $u_i$ now consists of an
index $j$ of $A$ together with an index into the adjacency array of $u_{i, j}$
in $P_{i, j}$. Note that the array $A$ contains many entries for each micro
boundary vertex. By Lemma~\ref{lem:duplicates} this is asymptotically no
problem, as the number of entries in all arrays $A$ depends linearly on the
number of micro boundary vertices. The arrays $A$ can be built in sublinear
time and space by iterating over all micro pieces, and inside each micro piece
iterating over all micro-boundary vertices. We store the analogous data
structure for all mini boundary vertices.

We now must store the state of each iterator during the DFS. Construct an array
storing for each boundary vertex the state of the iterator over $u$'s
neighborhood together with $u$'s parent in the DFS tree and the color
visited/unvisited. This uses $O(\log n)$ bits per mini boundary vertex. For
each mini piece $P_i$ we construct arrays storing the respective information of
micro boundary vertices. As each vertex $u$ (identified via mini label $u_i$)
in $P_i$ has a degree of $O(r)$ and must have its parent in $P_i$ (identified
by some mini label $v_i \in [r]$), this information can be stored with
$O(\log{r})=O(\lglgn)$ bits per micro boundary vertex. All other vertices are
handled by the lookup table. The mentioned data structures use $o(n)$ bits.

The DFS naturally computes the palm tree, but it remains to show how to store
it such that afterward we can execute common queries such as iteration over all
children (i.e., incident tree edges), or iteration over all back edges that
start/end at a given vertex. We describe this for iteration over all children
of a vertex, the other queries are realized in the exact same fashion. For
micro non-boundary vertices these queries can directly be provided by the table
lookup, or by recursive structure.
We thus focus on the boundary vertices. For each mini boundary vertex $u$ we
maintain a list $L$ containing the ids of all mini pieces $P$ such that $u$ has
a child in $P$. Iteration over the children of $u$ can then be
expressed as an iteration over $L$, and for each $i \in L$ we output all
children of $u$ in $P_i$. Concretely this is done by outputting all children of
$u_i$ (the mini label of $u$ in $P_i$) as their respective mini labels, and
then translating them to their global labels. For micro boundary vertices,
analogous structures are built. The space analysis is analogous to the space
analysis of the previous paragraph, i.e., it uses $o(n)$ bits total.  We
summarize the results in the next lemma.

\begin{lemma}\label{lem:palm}
    Let $G$ be a separable graph given as a nested division $\mathcal{D}$.
    In $O(n/\plglg{n})=o(n)$ time and $o(n)$ bits we can execute a DFS on $\mathcal{D}$
    and store the resulting palm tree $T$ such that the following queries
    are available for any vertex $u$, in constant time (per element output).
    \begin{itemize}
        \item Iteration over the children $v$ of $u$ in $T$, ordered by $\dfs{v}$.
        \item Iteration over all back edges starting/ending at $u$ in $T$.
        \item Output the parent $u$.
    \end{itemize}
\end{lemma}

\subsection{Meta Information of a Depth-First Search}\label{ssec:meta}

During the computation of a DFS it is
common to store various values that relate to the traversal such
as the $s$-$t$-numbering
$\dfs{u}$ of a vertex $u$. Or more complex values such as the lowpoint $\lp{u}$ of a
vertex $u$, defined as the lowest numbered vertex reachable via a path starting
at $u$ that consists of zero or more tree edges and at most one back edge.
We can augment $\mathcal{D}$ with this information as follows.
We describe this first for the simplest value, the $s$-$t$-numbering $\dfs{u}$.
Assume that we have constructed the palm tree of Lemma~\ref{lem:palm}.

\subparagraph*{$\mathbf s$-$\mathbf t$ Numbering.} 
We begin with an observation regarding the $\dfs{u}$ of some non-boundary
vertex $u$ assigned to some piece $P$ (the following description applies both
for micro and mini pieces), and denote with $k$ the number of vertices of $P$. 
As any DFS must enter $P$ via an entry vertex $v$, which is either the root,
or a boundary vertex.
Thus, for every non-boundary vertex $u$ of $P$ there exists an entry vertex $v$
with $\dfs{v} < \dfs{u}$ such that $v$ is part of an entry-exit pair $(v, x)$
of $P$, and unless $x$ is a null-exit, then $v$ lies on the tree path from $v$
to $x$. If there are multiple such entry-exit pairs, choose a pair with
$\dfs{x}-\dfs{u}$ minimized, i.e., the first exit traversed by the DFS after
visiting $u$. Assume that such an exit vertex exists. Then we can observe that
$\dfs{x} - k < \dfs{u} < \dfs{x}$. If we have the special case of the null-exit
we can observe that $\dfs{v} < \dfs{u} < \dfs{v}+k$ as $v$ is part of some
subtree that lies fully in $P$ that is rooted at $u$.
What we have shown is that the $s$-$t$-numbering of a non-boundary vertex $u$ can
be computed with a small offset value $k' \in [-k,k]$ together with the
$s$-$t$-numbering of some entry or exit vertex. Thus, $\dfs{u} = k' + \dfs{v}$ with
$k'$ defined as the \emph{local offset} and $v$ the \emph{reference boundary}.
It is crucial that the local offset is bounded by the number of vertices in $P$.

For a mini boundary vertex $u$ we store $\dfs{u}$ explicitly in an array with
$\log n$ bits each. For a mini non-boundary vertex $u$, and its mini label
$u_i$ in a piece $P_i$, we store the local offset $k'$ together with its
reference boundary $v$ (stored as its mini label $v_i$). Storing these values
explicitly would use $O(n\log\log n)$ bits total, and as such we again employ a
recursive strategy. We apply the mentioned strategy only for vertices being
mini non-boundary vertices, that are also micro boundary vertices and we store
the local offset and reference boundary explicitly. For non-boundary vertices
$v_{i, j}$ of a micro piece $P_{i, j}$ we use the table lookup to compute both
the local offset together with the reference boundary on-the-fly. 
This general strategy can be used for any strongly local value, possibly in
combination with some additional structures required for more involved queries.
Simple values such as the depth of a vertex, or the number of descendants, can
be handled exactly the same way as the $s$-$t$-numbering. For two of the more
involved values, lowpoint and lowest-common ancestor, we describe the details
next.

\begin{figure}[htbp]
    \centering
    \begin{subfigure}[b]{0.25\textwidth}
        \centering
        \includegraphics[width=\textwidth]{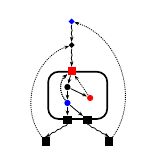}
    \end{subfigure}
    \begin{subfigure}[b]{0.25\textwidth}
        \centering
        \includegraphics[width=\textwidth]{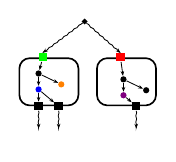}
    \end{subfigure}
    \caption{\textbf{Lowpoints (a)} The figure shows a piece and sketched parts of the DFS tree.
        Boundary vertices are drawn as squares, non-boundary vertices as
        circles, vertices of undefined type as diamonds. The lowpoint of the red non-boundary vertex is the red
        boundary vertex. The lowpoint of the blue non-boundary vertex is the
        blue diamond vertex. \textbf{LCAs (b)} The right figures shows two pieces. The LCA of the
        blue vertex and the violet vertex is the same as the LCA of the green
        and the red vertex. The LCA of the blue and the orange vertex is the
        green vertex. We can see that the LCA of two vertices in different pieces
        can be reduced to the LCA of two boundary vertices of the respective pieces.
}
    \label{fig:both}
\end{figure}

\subparagraph*{Lowpoints.}
The lowpoint $\lp{u}$ of a vertex $u$ is defined as the lowest $s$-$t$ number
$\dfs{v}$ such that $v$ is reachable via a path that traverses the DFS subtree
rooted at $u$ together with at most one back edge of the palm tree. The lowpoint
is a crucial value for typical DFS applications such as identifying biconnected
components. 
%
We show that $\lp{u}$ is strongly local. Assume that $u$ is a non-boundary
vertex in some piece $P$. Denote with $k$ the number of vertices in $P$. First
note that any back edge $(u, v)$ must have $v$ in $P$. Thus, $\lp{u}$ either is
$\dfs{u}$, depends on a back edge that lies within $P$, or it depends on the
lowpoint of a boundary vertex that is a descendant of $u$. Thus, we require
only the lowpoint of a boundary vertex of $P$, together with a small local
offset $x \in [k]$. See Figure~\ref{fig:both}a.

The standard algorithm due to Hopcroft and Tarjan computes the lowpoints during
a DFS traversal by updating it anytime the \textsc{DFS} backtracks to a vertex $u$ and
anytime a back edge $(u, v)$ is discovered. For non-boundary vertices inside a
micro piece, this update can be done via the table lookup using the
augmentations provided by the lookup table for each boundary vertex of a micro
piece $P_{i, j}$. Thus, anytime the DFS enters or leaves a micro piece $P_{i,
j}$, we can update the current lowpoint values of each non-boundary vertex of
$P_{i, j}$. For boundary vertices of a mini piece 
we use the same technique
as for the $s$-$t$-numbering, i.e., we store the required value in an array.
The runtime of updating all values across the DFS is $O(n/\plglg{n})$.

\subparagraph*{Lowest Common Ancestor.}
The operation $\lca{u, v}$ returns the root $w$ of the smallest subtree $T'$ of
a tree $T$ such that $u$ and $v$ are contained in $T$, i.e., it is the deepest
vertex $w$ in the tree such that $u$ and $v$ are descendants of $w$ where
deepest means it has the maximal distance from the root of $T$. 
We show that LCA queries for two vertices $u, v$ can be reduced to
strongly local values, one for each of the vertices.
Let $u, v$ be two non-boundary vertices of a piece $P$. If the LCA $w$ of $u,
v$ lies within $P$, then the value is clearly local. If the LCA $w$ does not
lie within $P$, there must be a path from $u$ to $w$ and $v$ to $w$ such that
each of these paths contains a boundary vertex $u'$ and $v'$, respectively. As
there can be multiple such boundary vertices, let it be the first ones
encountered by traversing the tree edges in reverse direction, i.e., the
nearest such vertices. Then the LCA of $u$ and $v$ is equal to the LCA of $u'$
and $v'$. Thus, for each pair of non-boundary vertices we only require to know
the $\lca{u',v'}$. The same is true, if $u$ and $v$ are non-boundary vertices
in different pieces. Again, $u'$ and $v'$ are the nearest ancestors being
boundary vertices in their respective pieces. Note that the choice of $u'$ and
$v'$ is fixed for $u$ and $v$, respectively, independent of the concrete query. 
See Figure~\ref{fig:both}b.
To handle the LCA between mini boundary vertices, and for each mini piece,
between micro boundary vertices, we use a known data structure to evaluate the
LCA queries
such as the data structure of Harel and Tarjan et al.~\cite{HarelT84}. For a
tree with $n'$ vertices, this data structure is initialized in $O(n')$ time and
uses $O(n' \log n)$ bits such that LCA queries are available in linear time.
We can not input our entire DFS tree to this data structure, and as such
construct special smaller trees that capture the ancestry only between mini boundary
vertices, and between micro boundary vertices for each mini piece.
The time to construct the necessary structures is $O(n/\plglg{n})$, and we use
$o(n)$ bits.

We describe the concrete construction of the required smaller trees below.
First observe the following. Let $T'$ be some tree with vertex set $V'$ and $S
\subset V$ a set of \emph{important vertices},
such that for pairs of vertices of $S$ we want to know their LCAs.
 First, all leaves $v$ in $V' \setminus S$ 
are irrelevant (can not be an LCA of a
pair of vertices in $S$) and 
we recursively can remove them. Next consider vertices of $V' \setminus S$
of degree $2$. These vertices are also irrelevant.
We can replace such vertex $v$ with its two edges $\{u,v\}\{v,w\}$ by a single edge
$\{u,w\}$. This tree
now contains the vertices of $S$ and all vertices that are LCA's of pairs of
vertices of $S$. Since we removed all vertices not in $S$ with degree $1$, 
the tree obtained has
$O(|S|)$ vertices.

Now consider the DFS tree $T$ and the set of mini boundary
vertices. Taking the set of mini boundary vertices as the important vertices,
we can see that we can shrink $T$ such that it only contains $O(n/\plg{n})$
vertices without losing any information regarding the LCAs between boundary
vertices. Analogously, for all micro boundary vertices inside a mini piece. The
shrunken tree $T_{P_{i, j}}$ for a micro piece $P_{i, j}$ 
can be built 
via table
lookup where we can support queries for every piece and every
arbitrary set of important vertices $S$ of the piece. Thus, we can obtain in $O(1)$
time a (set of) shrunken LCA tree(s) for piece $P_{i, j}$, which consists of
the shrunken whole DFS tree with respect to the boundary vertices of the
piece as important vertices.


To build the shrunken tree for all mini boundary vertices, we 
iterate over all mini pieces $P_{i, j}$ and obtain shrunken $T_{P_{i, j}}$
with respect to the mini boundary vertices as important vertices,
via table lookup.
We so can
build all required shrunken trees in $O(n/\plglg{n})$ time using $o(n)$
additional bits.

%
\begin{lemma}\label{lem:meta}
    Let $G$ be a separable graph given as a nested division $\mathcal{D}$
    such that $\mathcal{D}$ encodes the palm tree of some DFS in $\mathcal{D}$.
    After $O(n/\plglg{n})$ time preprocessing and using $o(n)$ additional bits
    we can provide the following queries.
    \ops{}
\end{lemma}

\subsection{Obtaining our Main Result}\label{ssec:result}

Theorem~\ref{th:main} follows from first constructing the encoding of
Lemma~\ref{lem:snd}, and then constructing the palm tree of
Lemma~\ref{lem:palm}. Lemma~\ref{lem:meta} provides the queries.
%
For planar graphs a compact variant of this encoding can be constructed such
that we only require linear time and bits during the construction step by
simply replacing the use of Lemma~\ref{lem:snd} with the use of
Lemma~\ref{lem:cnd}. 

Recall that the reason we can only provide a compact
encoding in linear time, is the use of Raman et al.'s FID~\cite{RamanRRS07} as an important
ingredient in the succinct variant. If one is fine with expected linear time
construction time, the encoding of the previous corollary remains succinct
instead of compact. For graphs that are not planar, the construction step
depends on how easily we can find separators. This shows
Corollary~\ref{cor:main}.

%
\bibliography{main}

\end{document}